\documentclass[prb,twocolumn,showpacs]{revtex4}
\usepackage{amsmath}
\usepackage{dcolumn}
\usepackage{graphicx}

\begin{document}

\title[Short title for running header]{The form and the origin of the orbital ordering in the electronic nematic phase of the Iron-based superconductors}
\author{Yuehua Su$^{1}$ Haijun Liao$^{2}$ and Tao Li$^{2}$}
\affiliation{$^{1}$Department of Physics, Yantai University, Yantai
264005, P.R.China\\ $^{2}$Department of Physics, Renmin University
of China, Beijing 100872, P.R.China}
\date{\today}

\begin{abstract}
We investigated the form of the orbital ordering in the electronic
nematic phase of the Iron-based superconductors by applying a group
theoretical analysis on a realistic five-band model. We find the
orbital order can be either of the inter-orbital s-wave form or the
intra-orbital d-wave form. From the comparison with existing ARPES
measurements of band splitting, we find the orbital ordering in the
122 system is dominated by an intra-orbital d-wave component, while
that in the 111 system is dominated by an inter-orbital s-wave
component. We find both forms of orbital order are strongly
entangled with the nematicity in the spin correlation of the system.
The condensation energy of the magnetic ordered phase is found to be
significantly improved(by more than 20 percents) when the degeneracy
between the $(\pi,0)$ and $(0,\pi)$ ordering pattern is lifted by
the orbital order. We argue there should be large difference in both
the scattering rate and the size of the possible pseudogap on the
electron pocket around the $\mathrm{X}=(\pi,0)$ and
$\mathrm{Y}=(0,\pi)$ point in the electronic nematic phase. We
propose this as a possible origin for the observed nematicity in
resistivity measurements.
\end{abstract}


\maketitle

The most exciting new feature of the Iron-based superconductors is
their multi-orbital nature. Many novel properties of the Iron-based
superconductors, especially those with the name of electronic
nematicity\cite{Chuang,Chu0,Chu1,Shen0,Chu2,Kasahara,Shimojima},
have been argued to be related to the orbital physics in these
systems\cite{Ku,Lv1,Singh,Lv2,Lee1,Nevidomskyy,Chubukov,Su}. The
ARPES observation of the rather large splitting between the
3$d_{xz}$ and 3$d_{yz}$-dominated band\cite{Shen0} in the magnetic
ordered phase further implies that the orbital degree of freedom is
deeply involved in the magnetic ordering phase transition. More
recent measurements find that the breaking of tetragonal symmetry in
the electronic structure can happen even without static magnetic
ordering\cite{Ding,Matsuda}. At the same time, the multi-orbital
character is also the key to understand the complex manifestation of
electron correlation effect and the structure of superconducting
pair in these systems\cite{Shen2,Kotliar}.

However, a comprehensive understanding on the role of orbital degree
of freedom in these systems is still in its infancy. In particular,
it is still a mystery how the observed electronic nematicity is
related to the orbital physics of the system. It is also unclear
what is the order parameter for the orbital ordering in the
electronic nematic phase of the Iron-based superconductors and how
it is entangled with the magnetic, structural and superconducting
properties of these systems.

The purpose of this paper is to determine the form of the order
parameter for the orbital ordering in the electronic nematic phase
of the Iron-based superconductors and explore its physical origin.
By applying group theoretical analysis to a realistic five band
model derived from band structure calculation, we are able to
determine the form of the order parameter for orbital ordering in
the electronic nematic phase of the Iron-based superconductors. We
find the orbital order can take either an inter-orbital s-wave form
or an intra-orbital d-wave form in the space spanned by the
degenerate $d_{xz}$ and $d_{yz}$ orbital. From the comparison with
ARPES measurements we find the orbital ordering in different
families of Iron-based superconductors take different forms. More
specifically, while the orbital order in the 122 system is dominated
by an intra-orbital d-wave component, the 111 system choose to order
mainly in the intra-orbital s-wave channel. We find both types of
orbital ordering are strongly entangled with the nematicity of spin
correlation in the system and can emerge spontaneously by lifting
the degeneracy between the magnetic ordering with wave vector
$(\pi,0)$ and $(0,\pi)$. We also argue that there should be large
difference in both the scattering rate and the size of pseudogap on
the electron pockets around $\mathrm{X}=(\pi,0)$ and
$\mathrm{Y}=(0,\pi)$ in the electronic nematic phase. This is
proposed to be the physical origin of nematicity observed in
resistivity measurements\cite{Chu0,Chu1,Chu2}.

To describe the complex electronic structure in the FeAs plane of
the Iron-based superconductors, we adopt the five-band tight binding
model derived from first principle calculation  of the LaFeAsO
system\cite{Kuroki}. The model reads,
\begin{eqnarray}
H_{band}=\sum_{i,j}\sum_{\mu,\nu,\sigma}t_{i,j}^{\mu,\nu}c_{i,\mu,\sigma}^{\dagger}c_{j,\nu,\sigma}.
\nonumber \label{eqn1}
\end{eqnarray}
Here $\mu,\nu=1,..,5$ is the index for the five maximally localized
Wannier functions(MLWFs) on the Fe site, namely,
$|1\rangle=|3d_{3\mathrm{Z}^{2}-\mathrm{R}^{2}}\rangle$,
$|2\rangle=|3d_{\mathrm{XZ}}\rangle$,
$|3\rangle=|3d_{\mathrm{YZ}}\rangle$,
$|4\rangle=|3d_{\mathrm{X}^{2}-\mathrm{Y}^{2}}\rangle$ and
$|5\rangle=|3d_{\mathrm{XY}}\rangle$. $t_{i,j}^{\mu,\nu}$ denotes
the hopping integral between the $\mu$-th and $\nu$-th orbital on
site $i$ and site $j$. The $\mathrm{X}$ and $\mathrm{Y}$-axis for
the Wannier functions are aligned with the Fe-As directions and are
rotated by 45 degrees from the $x$ and $y$-axis of the Fe-Fe square
lattice.

In the tetragonal phase, the point group around the Fe site is
$D_{2d}$, whose generators are $\sigma_{\mathrm{X}}$,
$\sigma_{\mathrm{Y}}$ and $R_{x}(\pi)$. Among the five MLWFs,
$|3\mathrm{Z}^{2}-\mathrm{R}^{2}\rangle$ belongs to the identity
representation, $|\mathrm{XY}\rangle$ and
$|\mathrm{X}^{2}-\mathrm{Y}^{2}\rangle$ belong to the
one-dimensional $\mathrm{B}_{1}$ and $\mathrm{B}_{2}$
representation, $|\mathrm{XZ}\rangle$ and $|\mathrm{YZ}\rangle$
belong to the two-dimensional representation of the $D_{2d}$ group.
If we denote the transformation of five MLWFs under the generators
$R_{m}=R_{x}(\pi),\sigma_{\mathrm{X}},\sigma_{\mathrm{Y}}$ as
$R_{m}|\mu\rangle=\sum_{\mu'}D_{\mu',\mu}(R_{m})|\mu'\rangle$, then
the hopping integrals in the tetragonal phase should satisfy the
following equations
\begin{eqnarray}
t_{i,j}^{\mu,\nu}=\sum_{\mu',\nu'}D_{\mu',\mu}(R_{m})D_{\nu',\nu}(R_{m})t_{i',j'}^{\mu',\nu'},
\end{eqnarray}
in which $i'=R_{m}^{-1}i$.

In the electronic nematic phase, the local symmetry around each Fe
site is reduced to $D_{2}$. In principle, such a symmetry breaking
may originate from either the electronic(for example, the charge,
spin or orbital degree of freedom) or the lattice degree of freedom.
Here we will focus on the symmetry breaking pattern in the orbital
space. As a result of this symmetry breaking, additional terms that
is forbidden by Eq.(1) can appear in $t_{i,j}^{\mu,\nu}$. These
symmetry breaking terms can be interpreted as the order parameter of
the orbital order in the electronic nematic phase and can be
detected from the band splitting in ARPES measurements. The general
form of such symmetry breaking perturbation can be largely
determined by group theoretical analysis. Here we will illustrate
such a procedure for the on-site symmetry breaking term for clarity.

The basic idea is to distill all the bilinear Hamiltonian terms in
the orbital space that is allowed by the symmetry group of the
orthogonal phase($D_{2}$) but is forbidden by the symmetry group of
the tetragonal phase($D_{2d}$). This can be done by operating the
projection operators of the identity representation for the $D_{2d}$
and $D_{2}$ group on an arbitrary initial bilinear Hamiltonian. The
$D_{2}$ group has four one dimensional irreducible representations.
Among the five MLWFs, $|3\mathrm{Z}^{2}-\mathrm{R}^{2}\rangle$ and
$|\mathrm{XY}\rangle$ both belong to the identity representation,
$|\mathrm{X}^{2}-\mathrm{Y}^{2}\rangle$ belongs to the
$\mathrm{B}_{1}$ representation, the linear combinations
$|\mathrm{XZ}\rangle\pm|\mathrm{YZ}\rangle$ belong to the
$\mathrm{B}_{2}$ and $\mathrm{B}_{3}$ representations and have the
$d_{xz}$ and $d_{yz}$ character. Thus the symmetry allowed on-site
Fermion bilinear terms in the orthogonal phase have the general form
of
\begin{eqnarray}
H_{2}&=&\sum_{i,\sigma}(\beta_{1}c^{\dagger}_{i,1,\sigma}c_{i,1,\sigma}+\beta_{2}c^{\dagger}_{i,5,\sigma}c_{i,5,\sigma}+\beta_{3}c^{\dagger}_{i,4,\sigma}c_{i,4,\sigma})\nonumber\\
&+&\beta_{4}\sum_{i,\sigma}(c^{\dagger}_{i,1,\sigma}c_{i,5,\sigma}+c^{\dagger}_{i,5,\sigma}c_{i,1,\sigma})\nonumber\\
&+&\beta_{5}\sum_{i,\sigma}(c^{\dagger}_{i,2,\sigma}+c^{\dagger}_{i,3,\sigma})(c_{i,2,\sigma}+c_{i,3,\sigma})\nonumber\\
&+&\beta_{6}\sum_{i,\sigma}(c^{\dagger}_{i,2,\sigma}-c^{\dagger}_{i,3,\sigma})(c_{i,2,\sigma}-c_{i,3,\sigma}).\label{eqn8}
\end{eqnarray}
However, it can be shown that the bilinear forms
$c^{\dagger}_{i,1,\sigma}c_{i,1,\sigma}$,
$c^{\dagger}_{i,4,\sigma}c_{i,4,\sigma}$,
$c^{\dagger}_{i,5,\sigma}c_{i,5,\sigma}$, and
$c^{\dagger}_{i,2,\sigma}c_{i,2,\sigma}
+c^{\dagger}_{i,3,\sigma}c_{i,3,\sigma}$ all belong to the identity
representation of $D_{2d}$ group. When these symmetric perturbations
are removed from Eq.(\ref{eqn8}), we get the symmetric breaking
perturbations in the orthogonal phase, which now takes the simple
form of
\begin{eqnarray}
\Delta
H&=&\lambda_{1}\sum_{i,\sigma}(c^{\dagger}_{i,2,\sigma}c_{i,3,\sigma}+c^{\dagger}_{i,3,\sigma}c_{i,2,\sigma})\nonumber\\
&+&\lambda_{2}\sum_{i,\sigma}(c^{\dagger}_{i,1,\sigma}c_{i,5,\sigma}+c^{\dagger}_{i,5,\sigma}c_{i,1,\sigma}).\nonumber
\end{eqnarray}
Here $\lambda_{1}=\beta_{5}-\beta_{6}$, $\lambda_{2}=\beta_{4}$. If
we further restrict our consideration to the subspace spanned by the
$d_{\mathrm{XZ}}$ and $d_{\mathrm{YZ}}$ orbital, which are the most
relevant degree of freedom in the electronic nematic phase
transition, the only allowable on-site symmetry breaking
perturbation would be
\begin{eqnarray}
\Delta
H&=&\lambda_{1}\sum_{i,\sigma}(c^{\dagger}_{i,2,\sigma}c_{i,3,\sigma}+c^{\dagger}_{i,3,\sigma}c_{i,2,\sigma}).
\end{eqnarray}

The above arguments can be easily generalized to determine the form
of the symmetry breaking perturbations on longer distances. The
necessity for such nonlocal terms can be seen clearly from the
strong momentum dependence of the band splitting observed in ARPES
measurements on 122 systems\cite{Shen0}. Following the same
procedures, we find there are in total 9 independent symmetry
breaking perturbations on nearest neighboring Fe-Fe bonds, which can
be classified into the extended s-wave, p-wave and d-wave channels,
\begin{eqnarray}
\Delta H=\Delta H_{s}+\Delta H_{p}+\Delta H_{d}.\nonumber
\end{eqnarray}
In the subspace spanned by the $d_{\mathrm{XZ}}$ and
$d_{\mathrm{YZ}}$ orbital, only an inter-orbital extended s-wave
term and an intra-orbital d-wave term are allowed. Combining these
terms with the on-site term found above, we find that the symmetry
breaking perturbation in the orthogonal phase is given by
\begin{eqnarray}
\Delta
H&=&\eta_{1}\sum_{i,\sigma}(c^{\dagger}_{i,2,\sigma}c_{i,3,\sigma}+c^{\dagger}_{i,3,\sigma}c_{i,2,\sigma})\nonumber\\
&+&\eta_{2}\sum_{i,\delta,\sigma}(c^{\dagger}_{i,2,\sigma}c_{i+\delta,3,\sigma}+c^{\dagger}_{i,3,\sigma}c_{i+\delta,2,\sigma})\nonumber\\
&+&\eta_{3}\sum_{i,\delta,\sigma}\mathrm{d}_{\delta}(c^{\dagger}_{i,2,\sigma}c_{i+\delta,2,\sigma}+c^{\dagger}_{i,3,\sigma}c_{i+\delta,3,\sigma}).
\end{eqnarray}
Here $\delta$ denotes the vectors connecting nearest neighboring Fe
sites and $\mathrm{d}_{\delta}$ is the d-wave form factor. In
principle, symmetry breaking terms on longer bonds can also be
determined in the same manner. However, from the comparison with the
ARPES measurements, we find it suffices to keep only the on-site and
the nearest neighboring terms. A list of all independent symmetry
breaking terms up to the next nearest neighboring bonds is given in
the appendix for reference.

The three parameters $\eta_{1,2,3}$ in Eq.(4) can be determined from
fitting the momentum dependence of the band splitting found in ARPES
measurements. In particular, they can be extracted from the band
splitting at the high symmetry momentum of $\Gamma=(0,0)$,
$\mathrm{X}=(\pi,0)$, $\mathrm{Y}=(0,\pi)$ and
$\mathrm{M}=(\pi,\pi)$. From group theoretical point of view, the
electronic state at these high symmetry momentums should form the
basis functions of irreducible representation of the $D_{2}$ group,
which is also the wave vector point group at these momentums in the
electronic nematic phase. Thus, the electronic state in the
$3d_{xz}$ and $3d_{yz}$-dominated bands should have pure $d_{xz}$
and $d_{yz}$ character at these momentums. The band splitting at
these momentums are given by
\begin{eqnarray}
\Delta E_{d_{xz}}(\Gamma)&=&\eta_{1}+4\eta_{2}, \ \Delta E_{d_{yz}}(\Gamma)=-\eta_{1}-4\eta_{2}\nonumber\\
\Delta E_{d_{xz}}(\mathrm{X})&=&\eta_{1}-4\eta_{3}, \ \Delta E_{d_{yz}}(\mathrm{X})=-\eta_{1}-4\eta_{3}\nonumber\\
\Delta E_{d_{xz}}(\mathrm{Y})&=&\eta_{1}+4\eta_{3}, \ \Delta E_{d_{yz}}(\mathrm{Y})=-\eta_{1}+4\eta_{3}\nonumber\\
\Delta E_{d_{xz}}(\mathrm{M})&=&\eta_{1}-4\eta_{2}, \ \Delta E_{d_{yz}}(\mathrm{M})=-\eta_{1}+4\eta_{2}\nonumber\\
\end{eqnarray}
Since the $3d_{xz}$ and $3d_{yz}$-dominated bands are far away from
the Fermi level around the $\mathrm{M}$ point, we will focus on the
band splitting along the $\Gamma-\mathrm{X}$ and $\Gamma-\mathrm{Y}$
directions in the following.

The momentum dependence of the band splitting induced by the three
types of orbital orders are plotted in Fig.1. To see the band
splitting more clearly, we overlay the dispersion along the
$\Gamma-\mathrm{Y}$ direction on that along the $\Gamma-\mathrm{X}$
direction. In the tetragonal phase, the dispersion of the
$3d_{yz}$-dominated band along the $\Gamma-\mathrm{X}$ direction
should be identical with that of the $3d_{xz}$-dominated band along
the $\Gamma-\mathrm{Y}$ direction, which are plotted as thin lines
in Fig.1 for reference. The band splitting caused by the on-site
orbital order is nonzero in the whole Brillouin zone and is only
weakly momentum dependent. For the extended s-wave orbital order,
the band splitting reaches its maximum(denoted as $\Delta_{max}$) at
the $\Gamma$ point and is exactly zero at the $\mathrm{X}$ and
$\mathrm{Y}$ points. For the d-wave orbital order, the band
splitting vanishes at the $\Gamma$ point and $\Delta_{max}$ is
reached at the $\mathrm{X}$ and $\mathrm{Y}$ points.

We now compare the theoretical predictions with the ARPES
measurements. In the Co-doped Ba122 system, the observed band
splitting in the $\Gamma-\mathrm{X}$ and $\Gamma-\mathrm{Y}$
direction exhibits strong momentum dependence and is very similar to
the d-wave from presented above. In particular, the band splitting
is negligible small around the $\Gamma$ point but is the most
evident at the $\mathrm{X}$ and $\mathrm{Y}$ points\cite{Shen0}. On
the other hand, in both the Na111 system or Li111 system, the
observed band splitting is clearly nonzero at the $\Gamma$ point and
the momentum dependence along the $\Gamma-\mathrm{X}$ and
$\Gamma-\mathrm{Y}$ direction is much less pronounced than that in
Co-doped Ba122 system\cite{Feng,Shen1,Ding}. Thus the order
parameter for orbital ordering in the 111 systems is more likely of
the on-site form. In principle, an extended s-wave component is also
possible in the 111 system. However, as we will show below, the
appearance of the extended-s-wave component is very unlikely from
energetic considerations. Thus, the form of orbital ordering in the
iron-based superconductors depends on the family of the material
studied. It is interesting to know if there is any generic reason
that the Co-doped Ba122 system and the Na111 system or Li111 system
choose different orbital ordering patterns.

\begin{figure}[h!]
\includegraphics[width=6cm,angle=0]{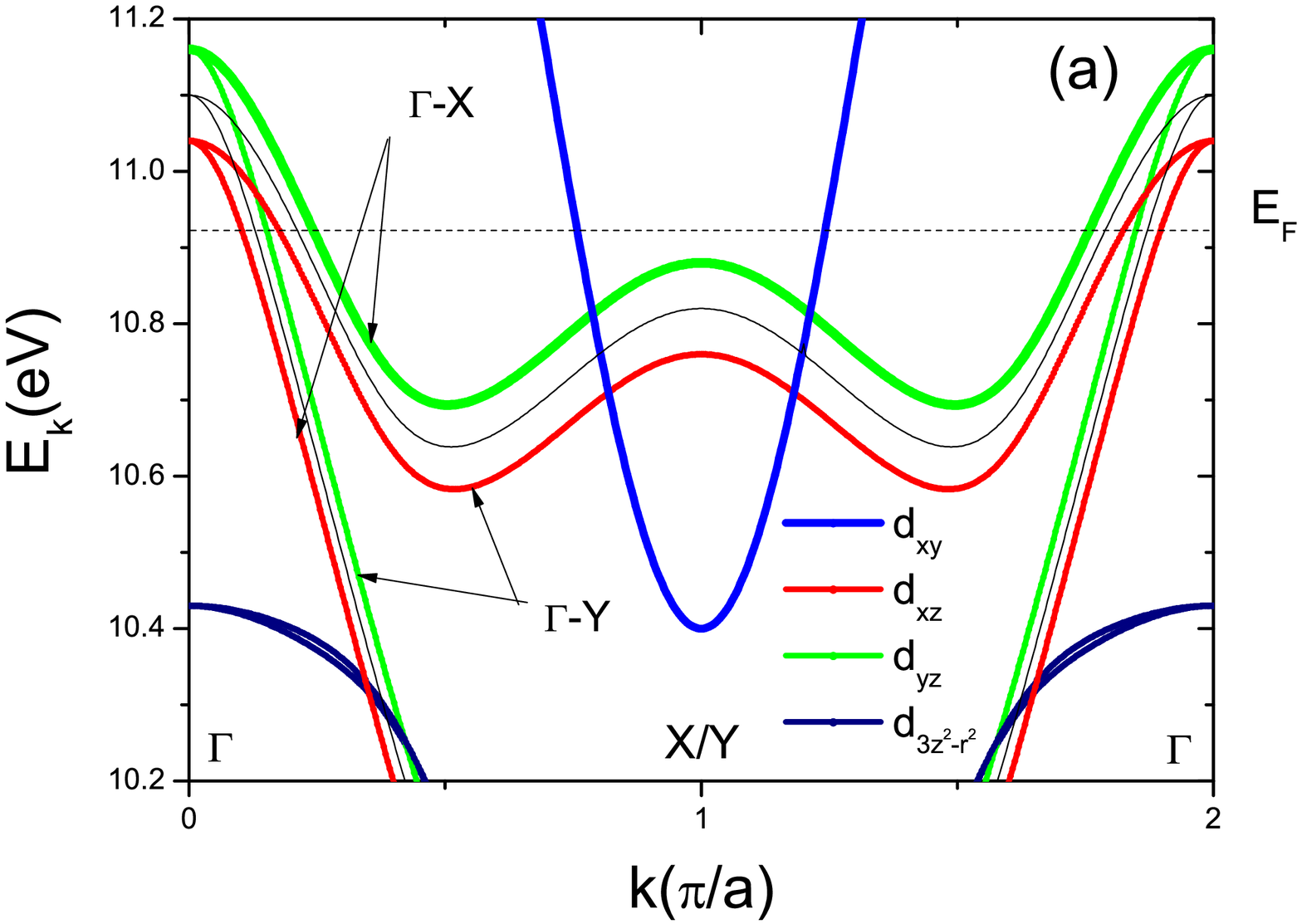}
\includegraphics[width=6cm,angle=0]{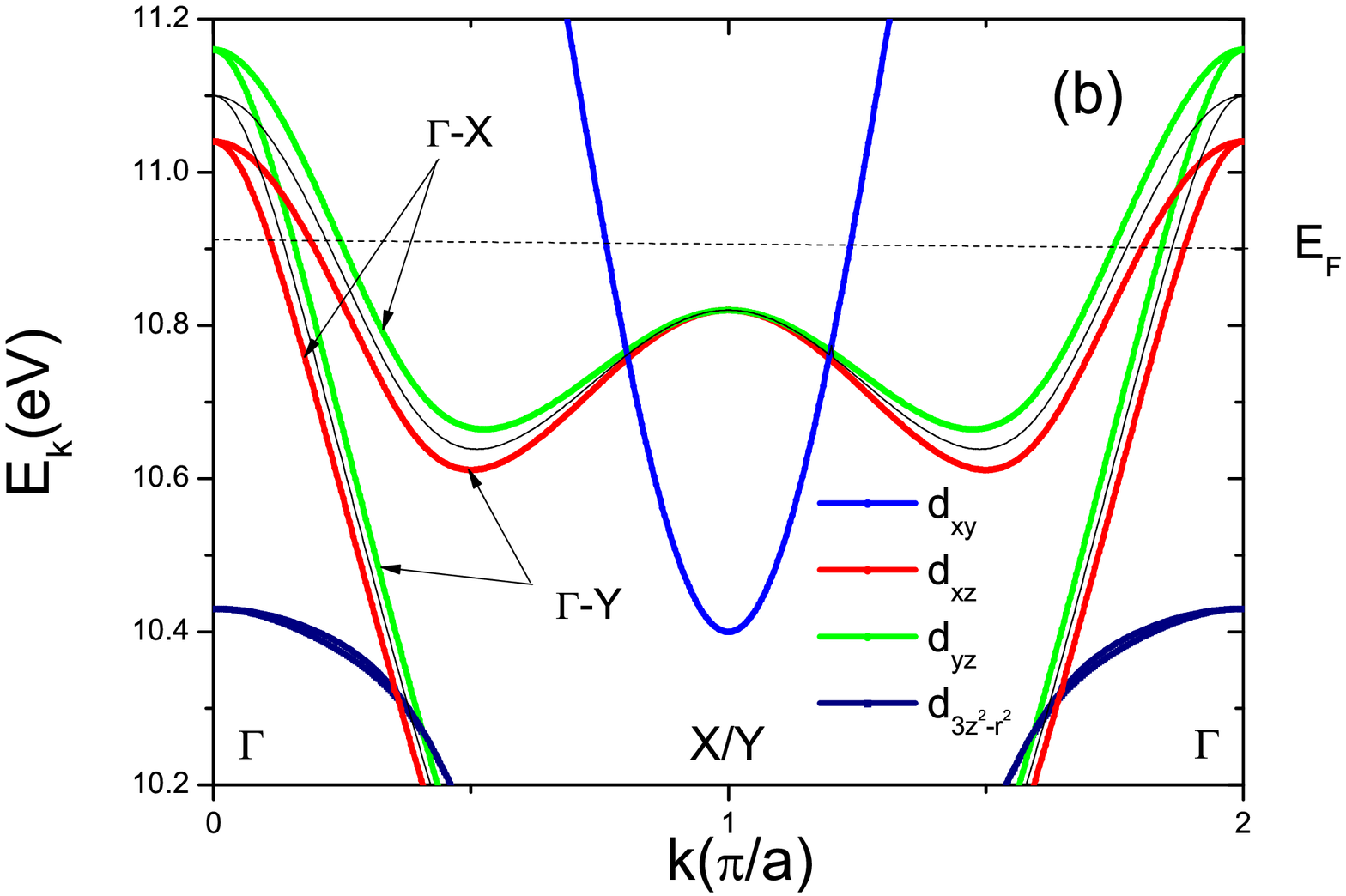}
\includegraphics[width=6cm,angle=0]{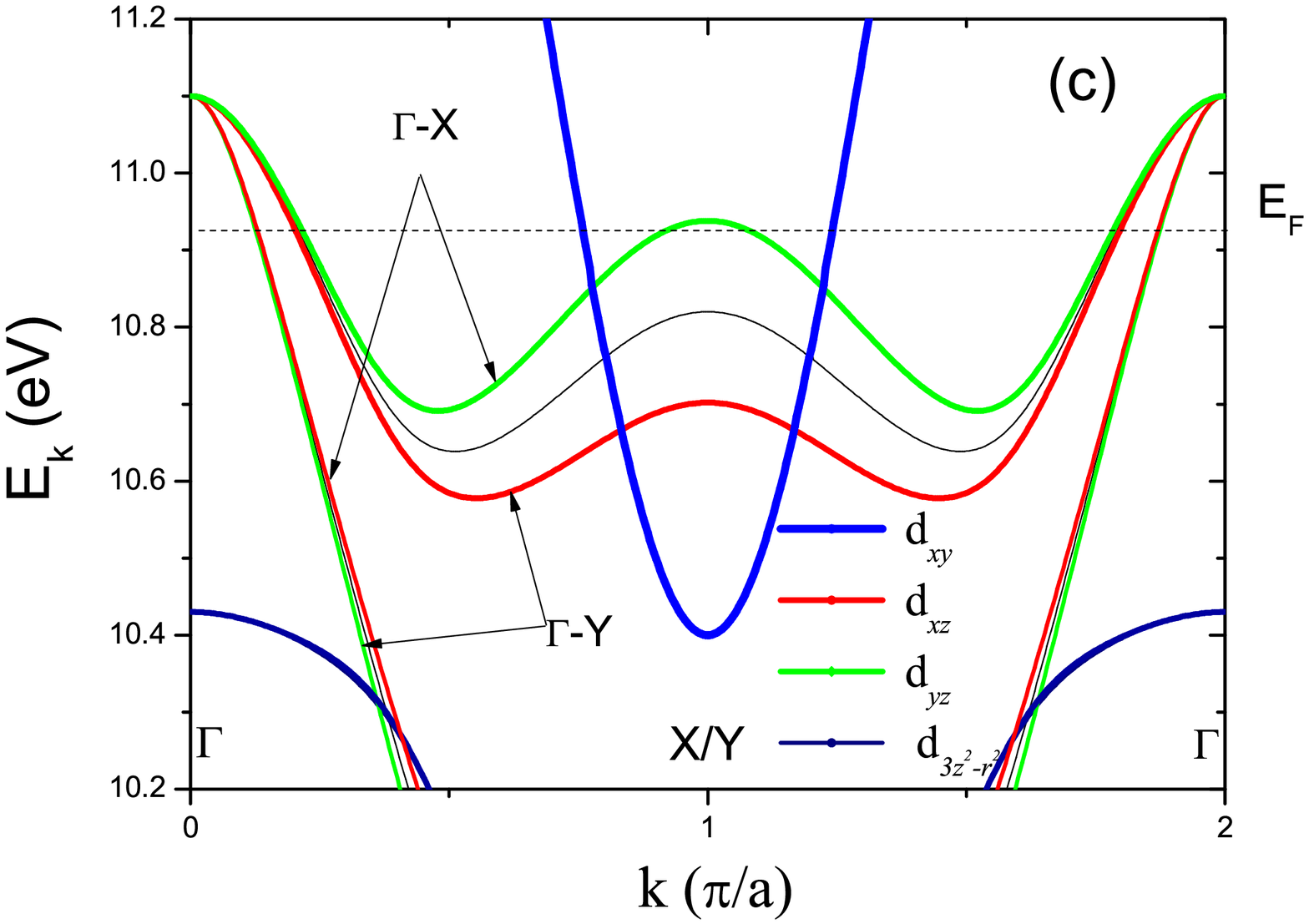}
\caption{Overlay of the band dispersion along the
$\Gamma-\mathrm{X}$ direction on that along the $\Gamma-\mathrm{Y}$
direction. The dominant orbital character is indicated by the color
of the lines and the dispersion in the tetragonal phase is plotted
in thin lines for reference. In the calculation we have set
$(\eta_{1},\eta_{2},\eta_{3})=(60,0,0)$ meV for the on-site case(a),
 $(0,15,0)$ meV for the extended s-wave case(b) and $(0,0,30)$ meV for the d-wave case(c). The dashed line
indicates the Fermi level at $n=6.1$. } \label{fig2}
\end{figure}

Now we discuss the possible origin of the orbital ordering in the
Iron-based superconductors. From the point of view of symmetry, the
orbital ordering in the electronic nematic phase can be just a
secondary effect caused by the breaking of tetragonal symmetry in
other channels such as the spin, charge or lattice degree of freedom
and has little contribution to the condensation energy of the
ordered phase. Here we will adopt a more exotic point of view and
assume that the orbital ordering can contribute significantly to the
condensation energy of the electronic nematic phase and thus emerge
spontaneously. This is a reasonable assumption since the size of the
observed band splitting in the electronic nematic phase is already
comparable to or even larger than the energy scales of other major
ordering tendencies in the system such as SDW order and
superconductivity. However, RPA calculations on the five band model
indicates that the Iron-based superconductors are far from a pure
orbital ordering instability. To resolve this problem, we assume
that the orbital ordering is strongly entangled with the nematicity
of spin correlation in the Iron-based superconductors, which is
dominated by two degenerate channels with wave vector $(\pi,0)$ and
$(0,\pi)$. As we will show below, the orbital order can lift such a
degeneracy and help the system to gain significantly more
condensation energy by selecting the favored ordering wave vector in
a way similar to the conventional Jahn-Teller effect.

\begin{figure}[h!]
\includegraphics[width=8cm,angle=0]{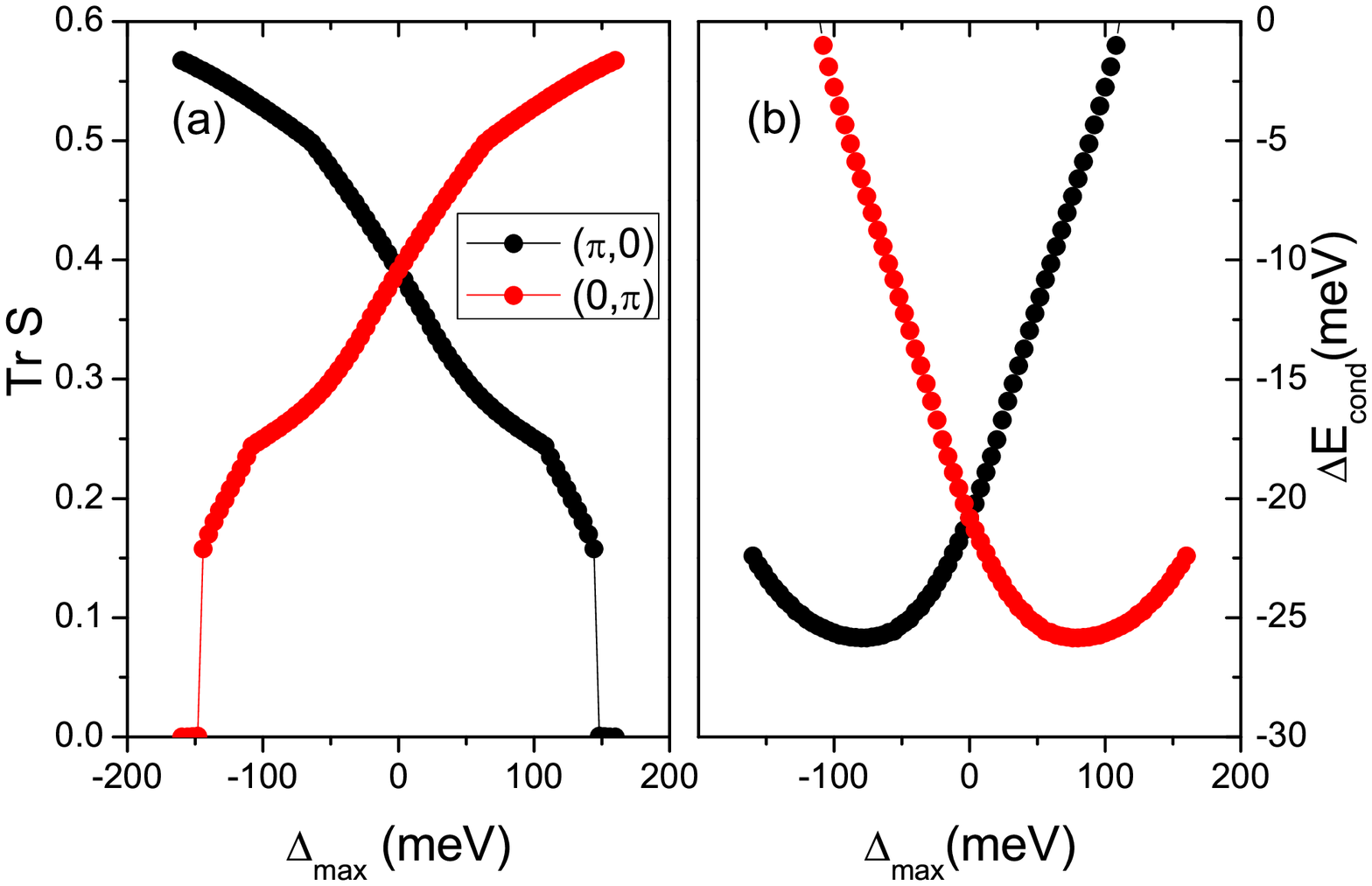}
\includegraphics[width=8cm,angle=0]{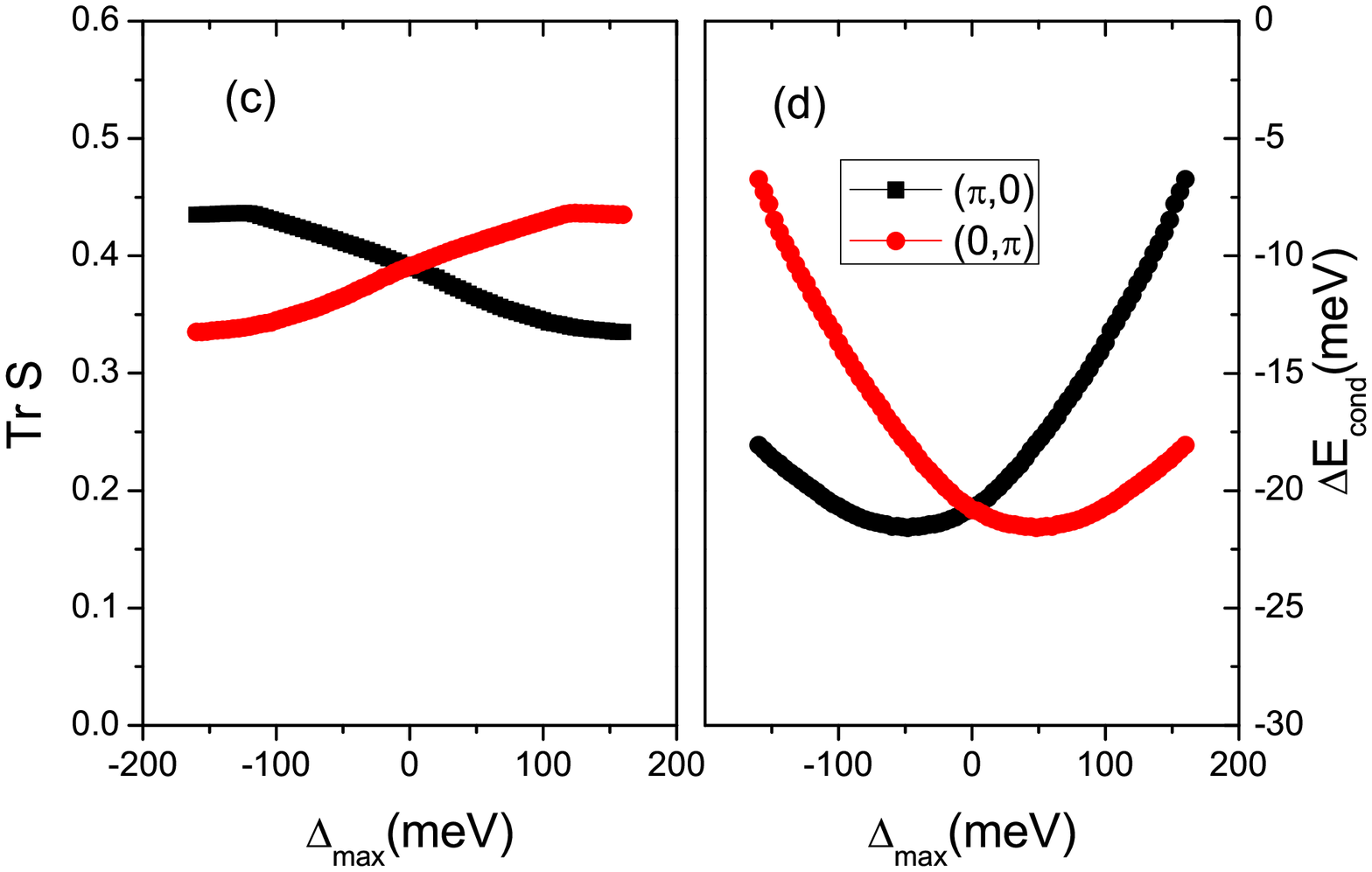}
\includegraphics[width=8cm,angle=0]{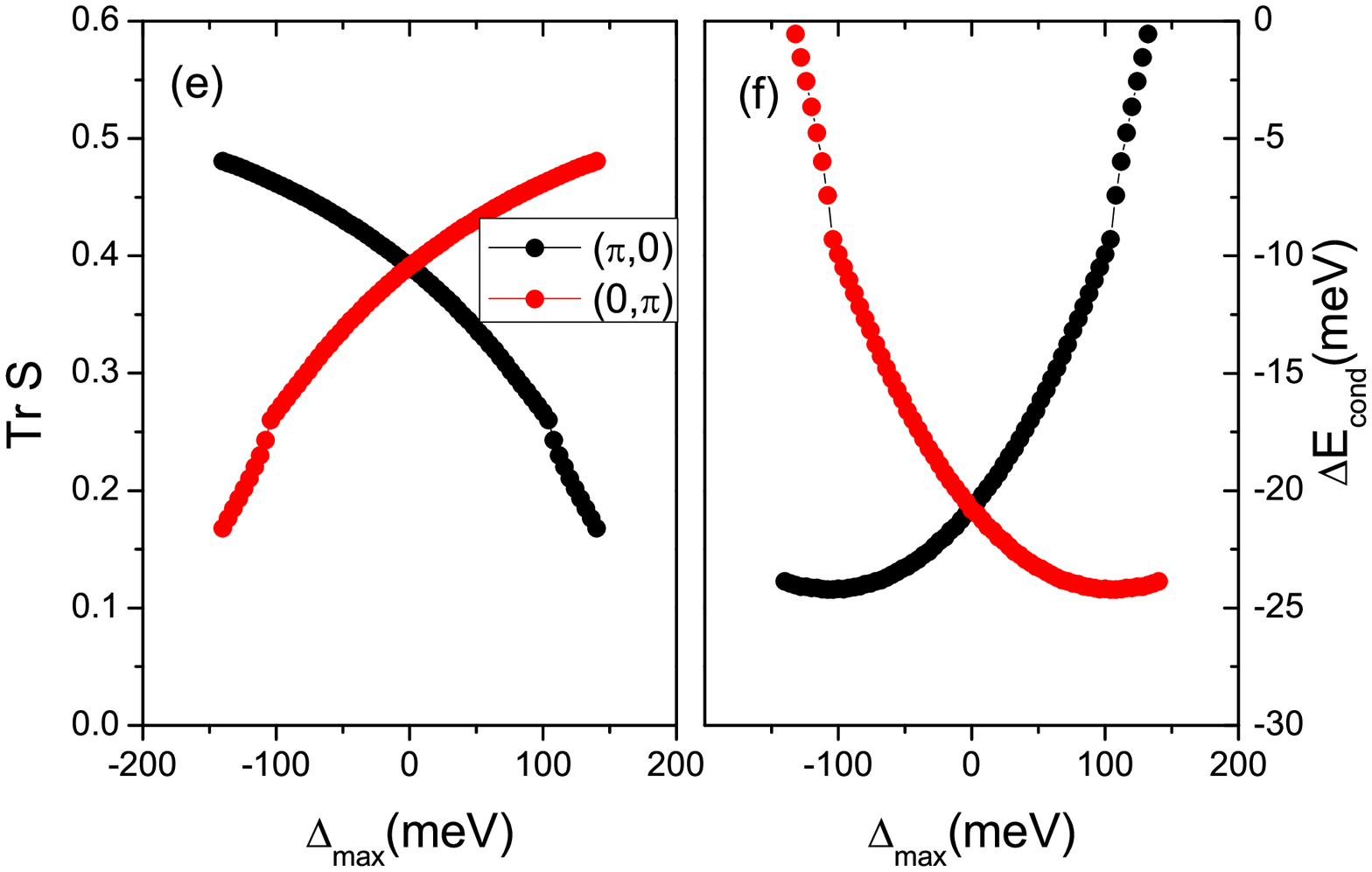}
\caption{Selection of SDW ordering pattern by orbital ordering.
Left: the ordered moment in the presence of on-site(a), extended
s-wave(c) and d-wave(e) orbital order. Right: the condensation
energy per unit cell in the presence of on-site(b), extended
s-wave(d) and d-wave(f) orbital order. } \label{fig2}
\end{figure}

In the following, we will illustrate the coupling between the
orbital order and the nematicity in the spin correlation at the mean
field level for the SDW ordered state. For this purpose, we
introduce the standard Kanamori-Hubbard Hamiltonian for the on-site
interactions on the Fe site. We then solve the mean filed equations
for the SDW order parameters in the presence of orbital order. The
SDW order is assumed to be collinear and has wave vector of either
$\mathrm{Q}_{x}=(\pi,0)$ or $\mathrm{Q}_{y}=(0,\pi)$. More
specifically, the SDW order parameter is assumed to be of the form
$\langle\frac{1}{2}\sum_{\sigma}\sigma
c^{\dagger}_{i,\mu,\sigma}c_{i,\nu,\sigma}\rangle=e^{i\mathrm{Q}_{x,y}\cdot
\mathrm{R}_{i}}S_{\mu,\nu}$, in which $S_{\mu,\nu}$ is a $5\times5$
matrix describing the distribution of the magnetic moment in the
orbital space. In the following, we will adopt $\mathrm{Tr}
S=\sum_{\mu}S_{\mu,\mu}$ as a measure of the magnitude of the
ordered moment. In the calculation, we set the interaction strength
as $U=1.5$eV, $U'=1.0$eV and $J_{H}=0.25$eV, which is large enough
to induce a moderate-sized ordered moment. The electron density is
fixed at $n=6.0$ in the calculation.

The solution to the mean field self-consistent equations is shown in
Fig.2. We find all the three types of orbital ordering can couple to
the nematicity in spin correlation. However, the strengthes of the
coupling are quite different for the three types of orbital
ordering. In particular, the coupling of the extended s-wave orbital
order to the nematicity of spin correlation is much weaker than that
of the on-site and the d-wave orbital order. For both the on-site
and the d-wave orbital order, the disfavored ordering pattern is
totally suppressed when the maximal band splitting $\Delta_{max}$
exceeds 160 meV, while in the case of the extended s-wave orbital
order, the change in the size of ordered moment is less than 20
percent even for $\Delta_{max}=200 meV$. To see if the orbital order
can emerge spontaneously from such a coupling, we calculate the
condensation energy of the system as a function of the orbital
order, which is also shown in Fig.2. We find both the on-site and
the d-wave orbital order can improve the condensation energy
significantly(more than 20 percents) and a sizeable orbital order
can be stabilized in both channels. On the other hand, the
improvement to the condensation energy from the extended s-wave
orbital order is rather small and the induced orbital order is also
much smaller than that in the other two channels.

Such a difference in the coupling strength can be understood
intuitively in the weak coupling picture from the nesting property
of the Fermi surface, which is important for the spin correlation.
More specifically, the band splitting between the $\mathrm{X}$ and
the $\mathrm{Y}$ point will enhance the nesting of the electron
pocket(with the hole pocket around the $\Gamma$ point) around one of
these two momentums and suppress the nesting of the electron pocket
around the other momentum. However, in the extended s-wave channel,
the band splitting is suppreseed around the $\mathrm{X}$ and the
$\mathrm{Y}$ point. This explains its weak entanglement with the
nematicity in the spin correlation. Since the orbital order can make
such a significant contribution to the condensation energy of the
ordered phase, it is more reasonable to think of the orbital order
as a component of a composite order parameter involving both the
spin and the orbital degree of freedom, rather than a secondary
effect of magnetic ordering.

We note the entanglement between the orbital order and the
nematicity in spin correlation is not limited in the magnetic
ordered phase. In the paramagnetic phase, such a coupling can be
realized through the so called Aslamazov-Larkin type vertex
correction\cite{Kontani}. A recent ARPES study of the
BaFe$_{2}$(AsP)$_{2}$ has found simultaneously the evidence of
pseudogap opening on the Fermi surface, which is most likely due to
strong scattering with spin fluctuation, and the band splitting
between the $\mathrm{X}$ and $\mathrm{Y}$ point in the electronic
nematic phase of the system\cite{Matsuda}. Following our line of
reasoning, we should expect rather different scattering rate and
different size of the pseudogap on the electron pocket around the
$\mathrm{X}$ and $\mathrm{Y}$ point in the electronic nematic phase
of the Iron-based superconductors. We propose this as a possible
origin for the observed nematicity in resistivity measurements in
the electronic nematic phase.

In conclusion, by applying group theoretical analysis to a realistic
five band model, we have determined the form of the orbital ordering
in the electronic nematic phase of the Iron-based superconductors.
We find the orbital ordering in these systems can be either of the
inter-orbital s-wave form or the intra-orbital d-wave form. From the
comparison with the ARPES observations, we find the orbital ordering
in the 122 systems is dominated by the intra-orbital d-wave
component, while that in the 111 systems is better described with an
on-site intra-orbital form. From a mean field calculation, we find
both types of orbital ordering are strongly entangled with the
nematicity of the spin correlation and can emerge spontaneously in
the model we have studied. We find the orbital order can contribute
significantly to the condensation energy of the magnetic ordered
phase and it is thus more reasonable to think of the orbital order
as a component of a composite order parameter involving both the
spin and the orbital degree of freedom, rather than a secondary
effect of magnetic ordering. We predict that both the scattering
rate and the size of pseudogap should be quite different on the
electron pocket around the $\mathrm{X}$ and $\mathrm{Y}$ point in
the electronic nematic phase of the Iron-based superconductors. We
propose this as a possible origin for the observed nematicity in
resistivity measurements in the electronic nematic phase.

Yuehua Su is support by NSFC Grant No. 10974167 and Tao Li is
supported by NSFC Grant No. 10774187, No. 11034012 and National
Basic Research Program of China No. 2010CB923004. We are grateful to
K. Kuroki for clarifying the phase convention used in
Ref.\onlinecite{Kuroki}.

\section{Appendix}
In this appendix, we present all possible symmetry breaking terms in
the electronic nematic phase in the full five dimensional orbital
space up to the next nearest neighboring bonds. Here we will assume
that both the translational and the time reversal symmetry is
respected in the electronic nematic phase. Following the procedure
as outlined in the main text, we find there are in total 9
independent symmetry breaking perturbations on nearest neighboring
Fe-Fe bonds and 6 independent symmetry breaking perturbations on the
next nearest neighboring bonds. The form of the symmetry breaking
terms on the nearest neighboring bonds are given by
\begin{eqnarray}
\Delta H=\Delta H_{s}+\Delta H_{p}+\Delta H_{d},\nonumber
\end{eqnarray}
in which
\begin{eqnarray}
\Delta H_{s}&=&\kappa_{1}\sum_{i,\delta,\sigma}(c^{\dagger}_{i,2,\sigma}c_{i+\delta,3,\sigma}+c^{\dagger}_{i,3,\sigma}c_{i+\delta,2,\sigma})\nonumber\\
&+&\kappa_{2}\sum_{i,\delta,\sigma}(c^{\dagger}_{i,1,\sigma}c_{i+\delta,5,\sigma}+c^{\dagger}_{i,5,\sigma}c_{i+\delta,1,\sigma})\nonumber
\end{eqnarray}
\begin{eqnarray}
\Delta H_{p}&=&\kappa_{3}\sum_{i,\delta,\sigma}(\mathrm{p}_{\delta}c^{\dagger}_{i,1,\sigma}c_{i+\delta,2,\sigma}+\mathrm{p}'_{\delta}c^{\dagger}_{i,1,\sigma}c_{i+\delta,3,\sigma})\nonumber\\
&+&\kappa_{3}\sum_{i,\delta,\sigma}(\mathrm{p}_{\delta}c^{\dagger}_{i,2,\sigma}c_{i+\delta,1,\sigma}+\mathrm{p}'_{\delta}c^{\dagger}_{i,3,\sigma}c_{i+\delta,1,\sigma})\nonumber\\
&+&\kappa_{4}\sum_{i,\delta,\sigma}(\mathrm{p}_{\delta}c^{\dagger}_{i,3,\sigma}c_{i+\delta,5,\sigma}+\mathrm{p}'_{\delta}c^{\dagger}_{i,2,\sigma}c_{i+\delta,5,\sigma})\nonumber\\
&+&\kappa_{4}\sum_{i,\delta,\sigma}(\mathrm{p}_{\delta}c^{\dagger}_{i,5,\sigma}c_{i+\delta,3,\sigma}+\mathrm{p}'_{\delta}c^{\dagger}_{i,5,\sigma}c_{i+\delta,2,\sigma})\nonumber\\
&+&\kappa_{5}\sum_{i,\delta,\sigma}(\mathrm{p}_{\delta}c^{\dagger}_{i,2,\sigma}c_{i+\delta,4,\sigma}-\mathrm{p}'_{\delta}c^{\dagger}_{i,3,\sigma}c_{i+\delta,4,\sigma})\nonumber\\
&+&\kappa_{5}\sum_{i,\delta,\sigma}(\mathrm{p}_{\delta}c^{\dagger}_{i,4,\sigma}c_{i+\delta,2,\sigma}-\mathrm{p}'_{\delta}c^{\dagger}_{i,4,\sigma}c_{i+\delta,3,\sigma}),\nonumber
\end{eqnarray}
and
\begin{eqnarray}
\Delta
H_{d}&=&\kappa_{6}\sum_{i,\delta,\sigma}\mathrm{d}_{\delta}(c^{\dagger}_{i,2,\sigma}c_{i+\delta,2,\sigma}+c^{\dagger}_{i,3,\sigma}c_{i+\delta,3,\sigma})\nonumber\\
&+&\kappa_{7}\sum_{i,\delta,\sigma}\mathrm{d}_{\delta}c^{\dagger}_{i,1,\sigma}c_{i+\delta,1,\sigma}\nonumber\\
&+&\kappa_{8}\sum_{i,\delta,\sigma}\mathrm{d}_{\delta}c^{\dagger}_{i,4,\sigma}c_{i+\delta,4,\sigma}\nonumber\\
&+&\kappa_{9}\sum_{i,\delta,\sigma}\mathrm{d}_{\delta}c^{\dagger}_{i,5,\sigma}c_{i+\delta,5,\sigma}\nonumber
\end{eqnarray}
represent the extended s-wave, p-wave and d-wave components of the
symmetry breaking perturbations. Here $\mathrm{p}_{\delta}$,
$\mathrm{p'}_{\delta}$ are p-wave form factors,
$\mathrm{d}_{\delta}$ is the d-wave form factor. The value of these
form factors are illustrated in Fig.3.

\begin{figure}[h!]
\includegraphics[width=8cm,angle=0]{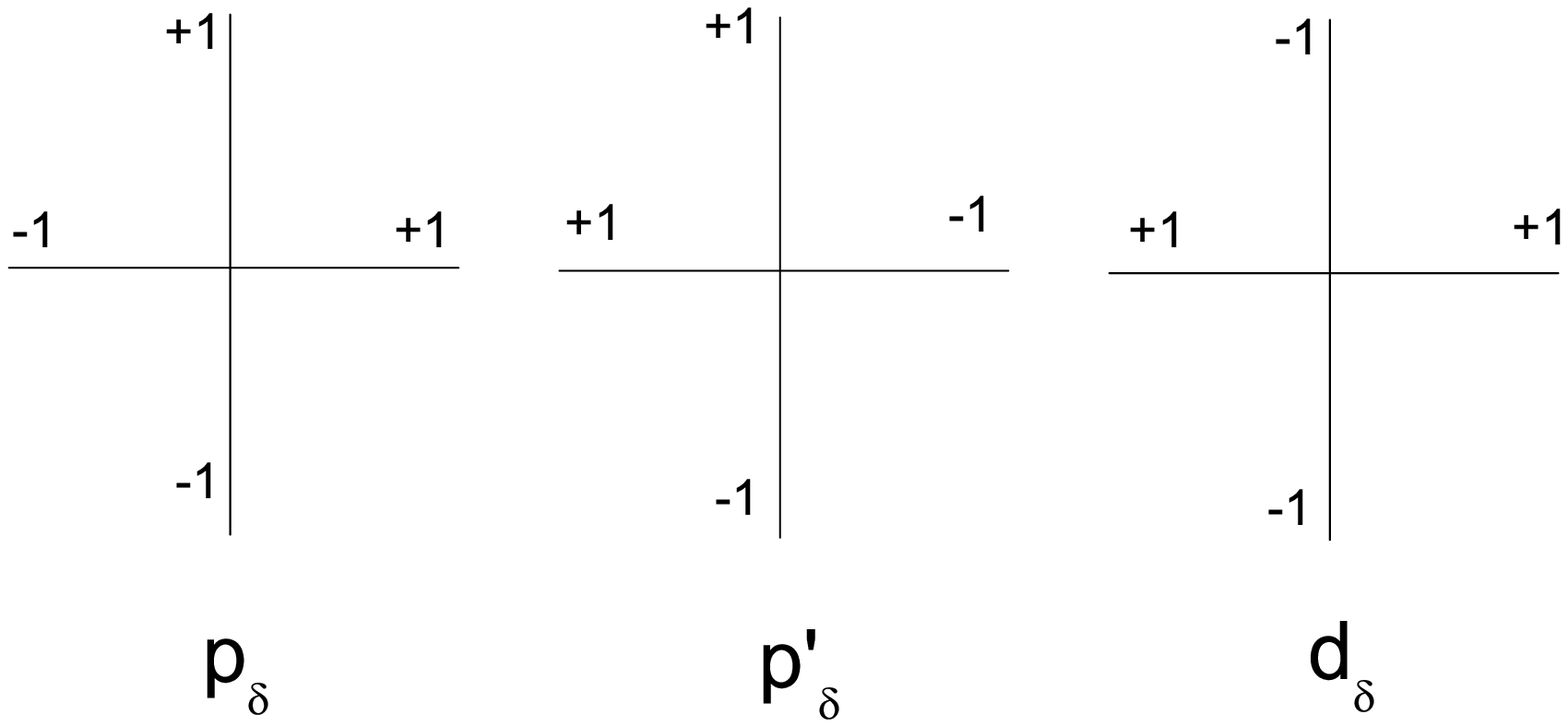}
\caption{An illustration of the p-wave and d-wave form factors on
the nearest neighboring bonds.}
\end{figure}

In the subspace spanned by the $d_{\mathrm{XZ}}$ and
$d_{\mathrm{YZ}}$ orbital, there are only two possible terms on the
nearest neighboring bond and are given by
\begin{eqnarray}
\Delta H&=&\eta_{2}\sum_{i,\delta,\sigma}(c^{\dagger}_{i,2,\sigma}c_{i+\delta,3,\sigma}+c^{\dagger}_{i,3,\sigma}c_{i+\delta,2,\sigma})\nonumber\\
&+&\eta_{3}\sum_{i,\delta,\sigma}\mathrm{d}_{\delta}(c^{\dagger}_{i,2,\sigma}c_{i+\delta,2,\sigma}+c^{\dagger}_{i,3,\sigma}c_{i+\delta,3,\sigma})
\end{eqnarray}
in which $\eta_{2}=\kappa_{1}$, $\eta_{3}=\kappa_{6}$.

The 6 allowed symmetry breaking perturbations on the next nearest
neighboring bonds are given by
\begin{eqnarray}
\Delta H'=\Delta H'_{s}+\Delta H'_{p}+\Delta H'_{d},\nonumber
\end{eqnarray}
in which
\begin{eqnarray}
\Delta H'_{s}&=&\kappa'_{1}\sum_{i,\delta',\sigma}(c^{\dagger}_{i,1,\sigma}c_{i+\delta',5,\sigma}+c^{\dagger}_{i,5,\sigma}c_{i+\delta',1,\sigma})\nonumber\\
&+&\kappa'_{2}\sum_{i,\delta',\sigma}(c^{\dagger}_{i,2,\sigma}c_{i+\delta',3,\sigma}+c^{\dagger}_{i,3,\sigma}c_{i+\delta',2,\sigma})\nonumber
\end{eqnarray}
\begin{eqnarray}
\Delta H'_{p}&=&\kappa'_{3}\sum_{i,\delta',\sigma}(\mathrm{p}_{\delta'}c^{\dagger}_{i,1,\sigma}c_{i+\delta',2,\sigma}+\mathrm{p}'_{\delta'}c^{\dagger}_{i,1,\sigma}c_{i+\delta',3,\sigma})\nonumber\\
&+&\kappa'_{3}\sum_{i,\delta',\sigma}(\mathrm{p}_{\delta'}c^{\dagger}_{i,2,\sigma}c_{i+\delta',1,\sigma}+\mathrm{p}'_{\delta'}c^{\dagger}_{i,3,\sigma}c_{i+\delta',1,\sigma})\nonumber\\
&+&\kappa'_{4}\sum_{i,\delta',\sigma}(\mathrm{p}_{\delta'}c^{\dagger}_{i,3,\sigma}c_{i+\delta',5,\sigma}+\mathrm{p}'_{\delta'}c^{\dagger}_{i,2,\sigma}c_{i+\delta',5,\sigma})\nonumber\\
&+&\kappa'_{4}\sum_{i,\delta',\sigma}(\mathrm{p}_{\delta'}c^{\dagger}_{i,5,\sigma}c_{i+\delta',3,\sigma}+\mathrm{p}'_{\delta'}c^{\dagger}_{i,5,\sigma}c_{i+\delta',2,\sigma})\nonumber\\
&+&\kappa'_{5}\sum_{i,\delta',\sigma}(\mathrm{p}_{\delta'}c^{\dagger}_{i,2,\sigma}c_{i+\delta',4,\sigma}-\mathrm{p}'_{\delta'}c^{\dagger}_{i,3,\sigma}c_{i+\delta',4,\sigma}),\nonumber\\
&+&\kappa'_{5}\sum_{i,\delta',\sigma}(\mathrm{p}_{\delta'}c^{\dagger}_{i,4,\sigma}c_{i+\delta',2,\sigma}-\mathrm{p}'_{\delta'}c^{\dagger}_{i,4,\sigma}c_{i+\delta',3,\sigma})\nonumber
\end{eqnarray}
and
\begin{eqnarray}
\Delta
H'_{d}&=&\kappa'_{6}\sum_{i,\delta',\sigma}\mathrm{d}_{\delta'}(c^{\dagger}_{i,4,\sigma}c_{i+\delta',5,\sigma}+c^{\dagger}_{i,5,\sigma}c_{i+\delta',4,\sigma}).\nonumber\\
\end{eqnarray}
Here $\delta'$ denote the vectors connecting next nearest
neighboring Fe sites and $\mathrm{p}_{\delta'}$,
$\mathrm{p}'_{\delta'}$  and  $\mathrm{d}_{\delta'}$ are the p-wave
and the d-wave form factor on next nearest neighboring bonds, which
are illustrated in Fig.4. In the subspace spanned by the
$d_{\mathrm{XZ}}$ and $d_{\mathrm{YZ}}$ orbital, there is only one
allowable symmetry breaking perturbation on the next nearest
neighboring bonds, which has the form of
\begin{eqnarray}
\Delta
H&=&\eta_{4}\sum_{i,\delta',\sigma}(c^{\dagger}_{i,2,\sigma}c_{i+\delta',3,\sigma}+c^{\dagger}_{i,3,\sigma}c_{i+\delta',2,\sigma}).\nonumber
\end{eqnarray}
Here $\eta_{4}=\kappa'_{2}$.
\begin{figure}[h!]
\includegraphics[width=7cm,angle=0]{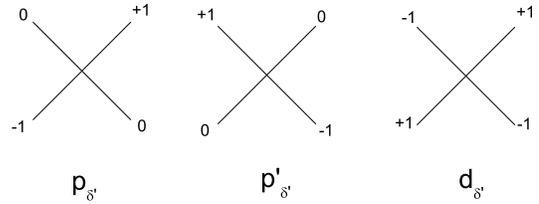}
\caption{An illustration of the p-wave and d-wave form factors on
the next nearest neighboring bonds.}
\end{figure}

\end{document}